\documentclass [12pt] {article}
\usepackage{graphicx}
\usepackage{epsfig}
\begin {document}
\begin{center}

{\Large {\bf PRODUCTION OF SECONDARIES IN HIGH ENERGY HEAVY ION
COLLISIONS}} \\

\vskip 1.5 truecm

{\bf J. Dias de Deus$^1$ and Yu. M. Shabelski}\\

\vskip 0.5 truecm

Petersburg Nuclear Physics Institute, Gatchina, St.Petersburg,
Russia \\

\vskip 1. truecm

\end{center}

\vskip 1.5 truecm

\begin{center}
{\bf ABSTRACT}
\end{center}

In the framework of Quark-Gluon String Model we calculate the
inclusive spectra of secondaries produced in heavy ion collisions
at intermediate (CERN SPS) and at much higher (RHIC) energies.
We demonstrate that the mechanism of secondary production changed
drastically in the energy interval $\sqrt{s} = 20 - 60$ GeV and
that is in agreement with qualitative estimates of Glauber-Gribov
theory. The results of numerical calculations at intermediate energies
are in reasonable agreement with the data. At RHIC energies
numerically large inelastic screening corrections should be accounted
for in calculations.

\vskip 1cm


\vskip 3.5 truecm

E-mail: shabelsk@thd.pnpi.spb.ru

\vskip 3. truecm

1) CENTRA, Istituto Superior Tecnico, Lisbon, Portugal

\newpage

\section{\bf Introduction}

The Quark--Gluon String Model (QGSM) and the Dual Parton Model (DPM)
are based on the Dual Topological Unitarization (DTU) and describe
quite reasonably many features of high energy production processes,
including the inclusive spectra of different secondary hadrons, their
multiplicities and multiplicity distributions, etc., both in
hadron--nucleon and hadron--nucleus collisions \cite{KTM}--\cite{Sh1}.
High energy interactions are considered as proceeding via the exchange
of one or several Pomerons and all elastic and inelastic processes
result from cutting through or between Pomerons \cite{AGK}. Inclusive
spectra of hadrons are related to the corresponding fragmentation
functions of quarks and diquarks, which are constructed using the
Reggeon counting rules \cite{Kai}.

In the case of interaction with nuclear target the Multiple Scattering
Theory (Gribov-Glauber Theory) is used and it allows to consider the
interaction with nuclear target as the superposition of interactions
with different numbers of target nucleons.

In the case of heavy ion collisions the Multiple Scattering Theory
also allows to consider this interaction as the superposition of
separate nucleon--nucleon interactions. However, in this case there
is no possibility to sum up all the diagrams in a rather simple form.
The first simple classes of diagrams can be accounted as the simple
expressions \cite{CM,Alkh}. The situation with more complicate
diagrams is not so clear \cite{Andr}--\cite{BoKa}.

In this paper we present the QGSM results for the calculation of the
inclusive spectra of secondaries produced in heavy ion collisions both
at intermediate (CERN SPS, $\sqrt{s_{NN}} = 17$ GeV) and much higher
(RHIC, $\sqrt{s_{NN}} = 60-200$ GeV) energies. The data at GSI and
CERN SPS energies \cite{E866}-\cite{Kaba} are in reasonable agreement
with Multiple Scattering Theory and with QGSM (see more detailed
discussion in Section 3). However the similar calculations overestimate
the data at RHIC energies by about a factor of two, as shown in Section
4. This effect was explained in \cite{CKTr} as large contribution of
inelastic screening correction namely in the case of heavy ion
collisions. We estimate the energy region where these correction
increase from several percents to twice as $\sqrt{s_{NN}}$ =
20 -- 60 GeV.

\section{\bf Inclusive spectra of secondary hadrons in the \newline
Quark-Gluon String Model}

For the quantitative predictions we need a model for multiparticle
production and we will use the QGSM for the numerical calculations
presented below.

As mentioned above, high energy hadron--nucleon and hadron--nucleus
interactions are considered in the QGSM as proceeding via the
exchange of one or several Pomerons. Each Pomeron corresponds to a
cylinder diagram, see Fig.~1a, and thus, when cutting a Pomeron,
two showers of secondaries are produced, as it is shown in Fig.~1b.
The inclusive spectrum of secondaries is determined by the convolution
of diquark, valence quark and sea quark distributions $u(x,n)$ in the
incident particles and the fragmentation functions $G(z)$ of quarks
and diquarks into secondary hadrons. The diquark and quark distribution
functions depend on the number $n$ of cut Pomerons in the considered
diagram.

\begin{figure}[htb]
\centering
\includegraphics[width=.5\hsize]{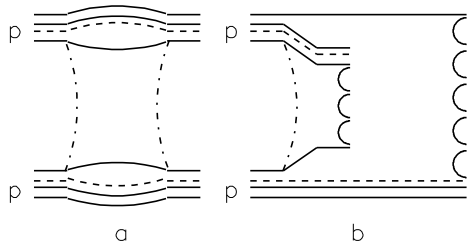}
\vskip -.5cm \caption{Cylinder diagram (cylinder is shown by
dash-dotted curves) corresponding to the one--Pomeron exchange
contribution to elastic $pp$ scattering (a) and its cut which determines
the contribution to inelastic $pp$ cross section (b). Quarks are shown
by solid curves and string junctions \cite{RV} by dashed curves.}
\end{figure}

In the case of a nucleon target the inclusive spectrum of a
secondary hadron $h$ has the form \cite{KTM}:

\begin{equation}
\frac{x}{\sigma_{inel}} \frac{d\sigma}{dx}
=\sum_{n=1}^{\infty}w_{n}\phi_{n}^{h}(x)\ \ ,
\end{equation}
where the functions $\phi_{n}^{h}(x)$ determine the contribution of
diagrams with $n$ cut Pomerons and $w_{n}$ is the probability of this
process \cite{TM}. Here we neglect the contributions of diffraction
dissociation processes which are comparatively small in most of
the processes considered below. This diffraction dissociation
contribution is important mainly for secondary production in large
$x_F$ region, which in not important in the present calculations.

For $pp$ collisions

\begin{equation}
\phi_{pp}^{h}(x) = f_{qq}^{h}(x_{+},n)f_{q}^{h}(x_{-},n) +
f_{q}^{h}(x_{+},n)f_{qq}^{h}(x_{-},n) +
2(n-1)f_{s}^{h}(x_{+},n)f_{s}^{h}(x_{-},n)\ \  ,
\end{equation}

\begin{equation}
x_{\pm} = \frac{1}{2}[\sqrt{4m_{T}^{2}/s+x^{2}}\pm{x}]\ \ ,
\end{equation}
where $f_{qq}$, $f_{q}$ and $f_{s}$ correspond to the contributions
of diquarks, valence quarks and sea quarks, respectively.

They are determined by the convolution of the diquark and quark
distributions with the fragmentation functions, for example,

\begin{equation}
f_{q}^{h}(x_{+},n) = \int_{x_{+}}^{1}
u_{q}(x_{1},n)G_{q}^{h}(x_{+}/x_{1}) dx_{1}\ \ .
\end{equation}

The diquark and quark distributions as well as the fragmentation
functions are determined from Regge intercepts \cite{Kai}.

In the case of nuclear targets we should consider the possibility
of one or several Pomeron cuts in each of the $\nu$ blobs of
hadron--nucleon inelastic interactions as well as cuts between
Pomerons. For example, for a $pA$ collision one of the cut
Pomerons links a diquark and a valence quark of the projectile proton
with a valence quark and diquark of one target nucleon. Other Pomerons
link the sea quark--antiquark pairs of the projective proton
with diquarks and valence quarks of another target nucleons and with
sea quark--antiquark pairs of the target.

For example, one of the diagram for inelastic interaction with two
target nucleons is shown in Fig.~2. In the blob of the $pN_1$
inelastic interaction one Pomeron is cut, and in the blob of $pN_2$
interaction two Pomerons are cut. It is essential to take into account
every possible Pomeron configuration and permutation in all digrams.
The process shown in Fig.~2 satisfies the condition
\cite{Sh2}--\cite{Jar} that the
absorptive parts of hadron--nucleus amplitude are determined by the
combinations of the absorptive parts of hadron--nucleon interactions.

\begin{figure}[htb]
\centering
\includegraphics[width=.5\hsize]{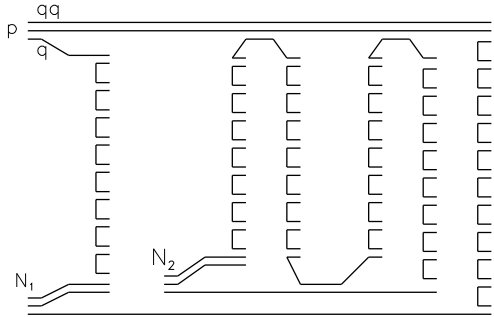}
\caption{One of the diagrams for inelastic interaction of an
incident proton with two target nucleons $N_1$ and $N_2$ in a $pA$
collision.}
\end{figure}

In the case of inelastic interactions with $\nu$ target nucleons
$n$ be the total number of cut Pomerons in $hA$ collisions ($n
\geq \nu$) and let $n_i$ be the number of cut Pomerons connecting
with the $i$-th target nucleon ($1 \leq n_i \leq n-\nu+1$). We
define the relative weight of the contribution with $n_i$ cut
Pomerons in every $hN$ blob as $w^{hN}_{n_i}$. For the inclusive
spectrum of the secondary hadron $h$ produced in a $p A$
collision we obtain \cite{KTMS}

\newpage

\begin{eqnarray}
\frac{x_E}{\sigma^{prod}_{pA}} \frac{d \sigma}{dx_F} & = &
\sum^A_{\nu=1} V^{(\nu)}_{pA} \left\{ \sum^{\infty}_{n=\nu}
\sum^{n-\nu+1}_{n_1 = 1} \cdot \cdot \cdot
\sum^{n-\nu+1}_{n_{\nu}=1} \prod^{\nu}_{l=1} w^{pN}_{n_l}
\right. \times \\ \nonumber & \times &
[f^h_{qq}(x_+,n)f^h_q(x_-,n_l) +
f^h_q(x_+,n)f^h_{qq}(x_-,n_l) + \\ \nonumber & + &
\sum^{2n-2}_{m=1} f^h_s(x_+,n)f^h_{qq,q,s}(x_-,n_m)] \left.
\right\} \;,
\end{eqnarray}
where $V^{(\nu)}_{pA}$ is the probability of "pure inelastic"
(nondiffractive) interactions with $\nu$ target nucleons, and we
should account for all possible Pomeron permutation and the
difference in quark content of the protons and neutrons in the
target.

In particular, the contribution of the diagram in Fig.~2 to the
inclusive spectrum is

\begin{eqnarray}
\frac{x_E}{\sigma^{prod}_{pA}} \frac{d \sigma}{dx_F} & = & 2
V^{(2)}_{pA} w^{pN_1}_1w^{pN_2}_2 \left\{
f^h_{qq}(x_+,3)f^h_q(x_-,1)\right. + \\ \nonumber & + &
f^h_q(x_+,3)f^h_{qq}(x_-,1) + f^h_s(x_+,3) [f^h_{qq}(x_-,2) +
f^h_q(x_-,2) + \\ \nonumber & + & 2f^h_s(x_-,2)] \left. \right\}
\;.
\end{eqnarray}

The diquark and quark distributions as well as the fragmentation
functions are here the same as in the case of a nucleon target.

\section{\bf Inclusive spectra in heavy ion collisions at
intermediate energies}

At comparatively low energies we do not have a simple model for the
calculation of the yields of secondaries. However, there exist the data
\cite{E866,E802}, which show that the yields of secondaries produced by
one interacting nucleon increase with the increase of the number of
interacting nucleons. An example is presented in Fig.~3 taken from
\cite{E802}.

\begin{figure}[htb]
\begin{center}
\mbox{\psfig{file=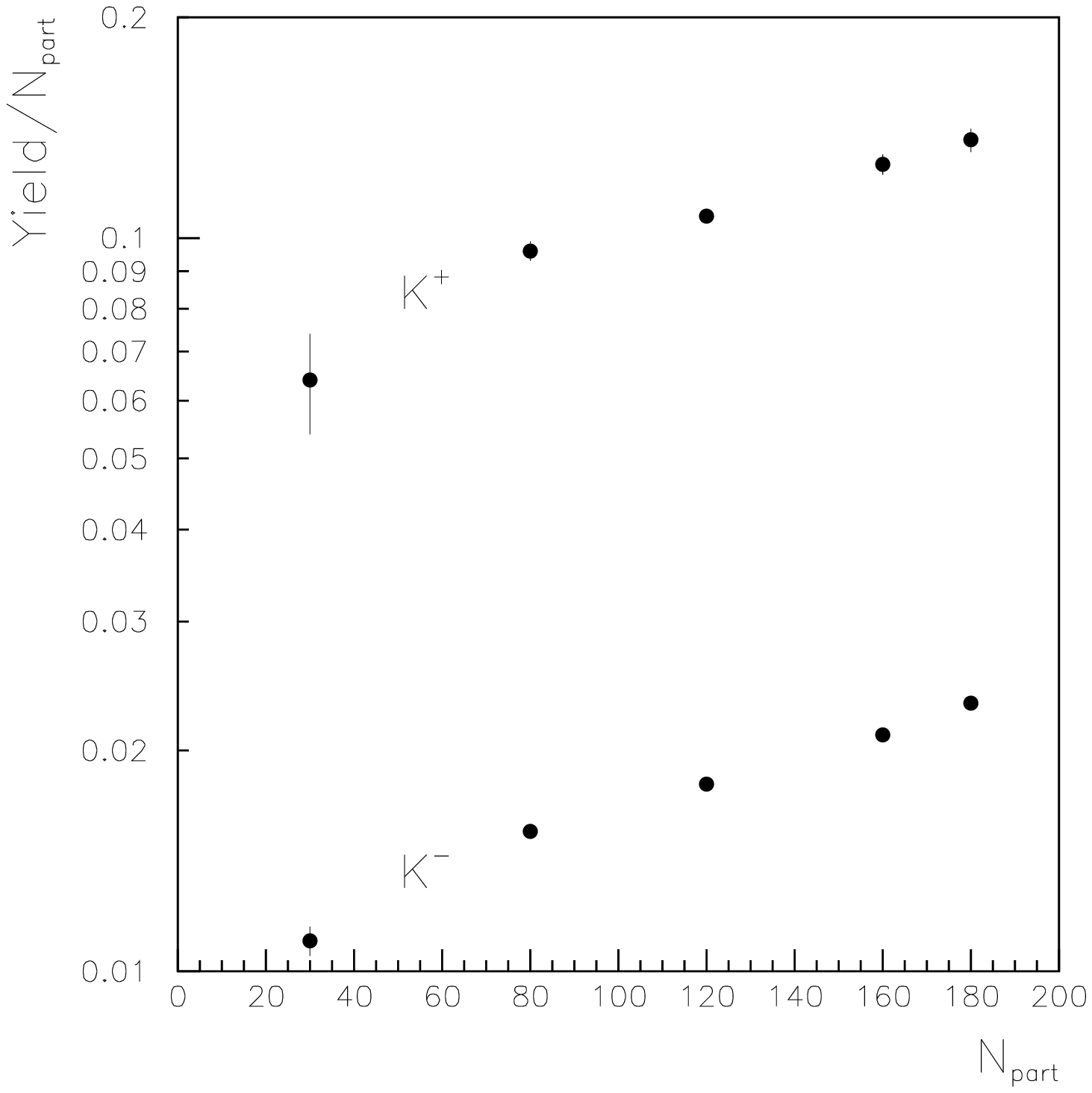,width=0.65\textwidth}} \\
Fig.~3. The total yields of kaons per projectile participant versus
the number of projectile participants $N_{part}$ in $Au-Au$ reactions
at $11.6A$ GeV/c.
\end{center}
\end{figure}

It is clear that the increase of the number of projectile participant,
$N_{part}$, means the decrease of the impact parameter and every
participant interacts, on the average, with larger number of target
nucleons, $\nu_{NA}$ \cite{PSh}. One can see from Fig.~3 that every
projectile nucleon produces more both $K^+$ and $K^-$ at small impact
parameters in comparison with peripheral interactions. These ratios
change about 2--2.5 time from large to small impact
parameters\footnote{Due to the energy correction factors \cite{Sh2}
these ratios should not be equal exactly to $\nu_{NA}$}. It means that
the multiplicity of the produced kaons is a real function of $\nu_{AB}$,
that is the number of elementary nucleon--nucleon collisions in the
case of inelastic interaction of nucleus A with nucleus B and
\begin{equation}
\langle \nu_{AB} \rangle = \frac{A B \sigma_{NN}^{inel}}
{\sigma_{AB}^{prod}} \;.
\end{equation}
This fact confirms the applicability of the Multiple Scattering Theory
at the energies of GSI, as a minimum.

The similar situation probably takes place at CERN SPS energies, see
Fig.~10 in \cite{Sik}. The saturation of the yields of secondaries
produced by one interacting nucleon possibly can be seen only at
$N_{pp} > 300$, i.e. in very central events. However, it is necessary
to note that the data \cite{Sik} are in some disagreement with the data
\cite{Kaba}.

At CERN SPS energies $158A$ GeV/c, following to the data \cite{Sik}, we
can use the QGSM and calculate the spectra of secondaries. In the case
of heavy ion collisions the energy conservation effects violate the
asymptotical ratio
\begin{equation}
\frac1{\sigma^{prod}_{AB}} \frac{d \sigma (AB \to hX)}{d y} =
\langle \nu_{AB} \rangle \frac1{\sigma^{inel}_{NN}}
\frac{d \sigma (NN \to hX)}{d y}
\end{equation}
significantly stronger than in the case of hadron--nucleus
collisions, where the asymptotical value is
\begin{equation}
\frac1{\sigma^{prod}_{NA}} \frac{d \sigma (NA \to hX)}{d y} =
\langle \nu_{NA} \rangle \frac1{\sigma^{in}_{NN}} \frac{d \sigma
(NN \to hX)}{d y}
\end{equation}
with
\begin{equation}
\langle \nu_{NA} \rangle = \frac{A \sigma_{NN}^{inel}}
{\sigma_{NA}^{prod}} \;.
\end{equation}

The possible way to account for the energy conservation effects was
suggested in \cite{Sh3}. The idea is that we will use the rigid target
approximation in two different ways. In the case of forward hemisphere
(nucleus $A$ fragmentation region) we account for that each nucleon
of the nucleus $A$ can interact with several nucleons of the nucleus
$B$, but the nucleons of $B$ interact not more than once. This is
equivalent to the case, where $A$ uncoupled nucleons, with the
corresponding impact parameter distribution, interact with the nucleus
$B$. In the backward hemisphere (nucleus $B$ fragmentation region) we
will use the same, but with change $A$ and $B$. Now $B$ uncoupled
nucleons interact with the nucleus $A$. The two contributions in the
central region are matched with good accuracy.

The inclusive spectrum of secondaries produced in $A-B$ collision can
be written as

\begin{eqnarray}
\frac1{\sigma^{prod}_{AB}} \frac{d \sigma (AB \to hX)}{d y} & = &
\theta (y) R^h_A(y) \langle N_A \rangle \frac1{\sigma^{prod}_{NB}}
\frac{d \sigma (NB \to hX)}{d y} + \;\;\; \\ \nonumber
& + &\theta (-y) R^h_B(-y) \langle N_B \rangle
\frac1{\sigma^{prod}_{NA}} \frac{d \sigma (NA \to hX)}{d y} \;,
\end{eqnarray}
where $y$ is the rapidity of secondary $h$ in c.m. frame and the
functions $R^h_{A,B}(y)$ account for the energy conservation effects.

Here we connect the inclusive spectra of secondaries in the heavy ion
collisions with the spectra in the nucleon--nucleus interactions. To
calculate the last ones, as well as the functions $R^h_{A,B}(y)$ we use
QGSM and the Multiple Scattering Theory for $NA$ collisions
\cite{KTMS}.

The functions $R^h_{A,B}(y)$ can be taken in the form \cite{Sh3}
\begin{equation}
R^h_A(y) = \frac{f^h(-y, \langle \nu \rangle_{NA})}
{\langle\nu \rangle_{NA} f^h(-y,1)} \;,
\end{equation}
where $f^h(y,\nu)$ is the contribution to the secondary $h$ spectrum
from a beam nucleon interaction with $\nu$ target nucleons, and
$f^h(y,1)$ is the spectrum of secondary particle $h$ in $NN$ collision.

The rapidity distributions of secondary $\pi^-$, $K^+$ and $K^-$
\cite{Afan}, as well as $\pi^+$, $p$ and $\bar{p}$ \cite{Sik}
measured in $Pb-Pb$ central collisions at 158 GeV/c per nucleon
are compared with our calculations using Eq.~(11) in Fig.~4.

\begin{figure}[htb]
\begin{center}
\vspace {.3cm}
\mbox{\psfig{file=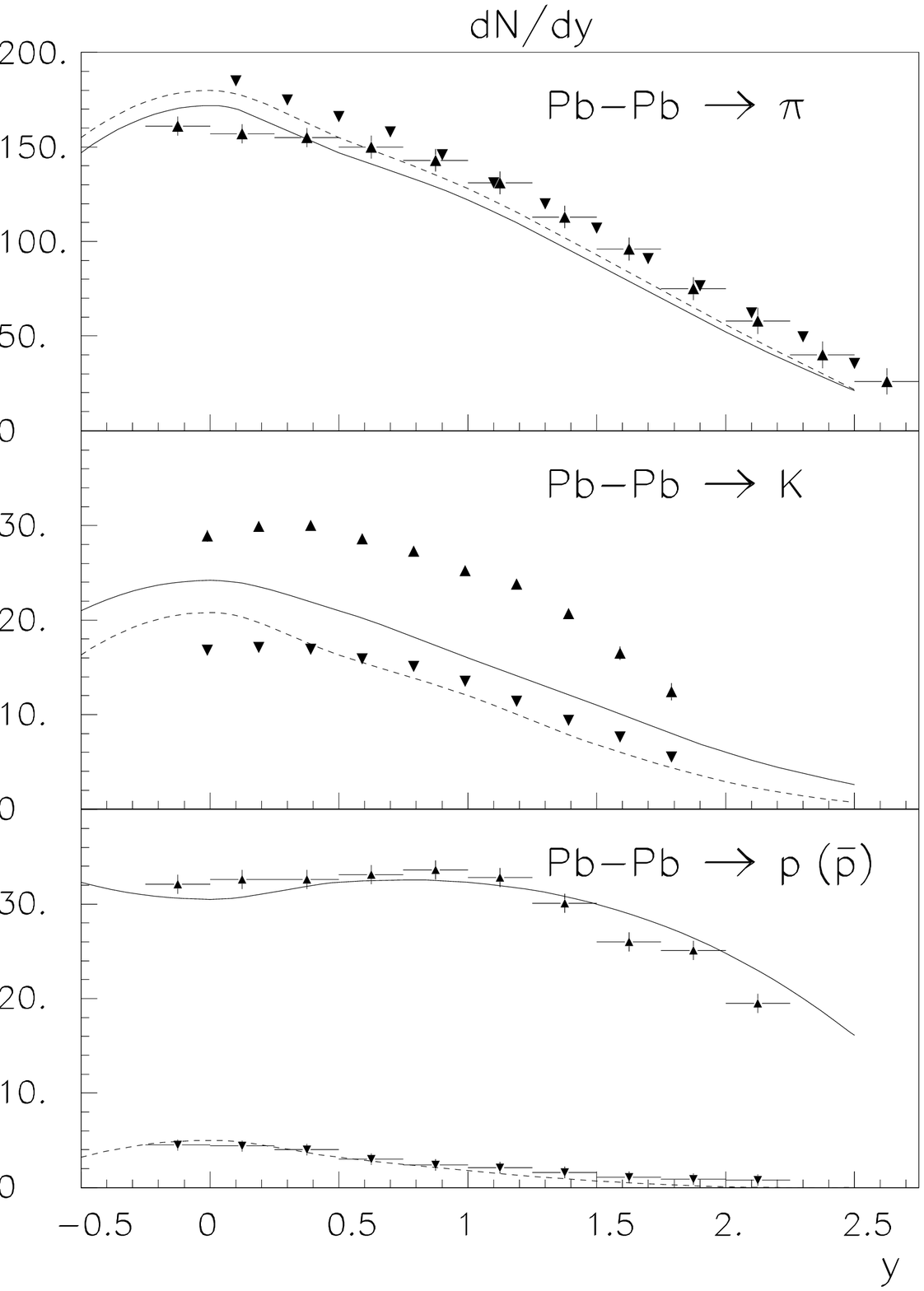,width=0.45\textwidth}} \\
Fig.~4. The rapidity distributions of secondary $\pi^+$,
$\pi^-$, $K^+$, $K^-$, $p$ and $\bar{p}$ produced in the central
$Pb-Pb$ collisions at 158 GeV/c per nucleon. Triangles show the
results for the positive secondaries and over turned triangles for
the negative ones. The solid curves are the QGSM predictions for
positive secondaries and dashed curves for the negative ones.
\end{center}
\end{figure}

In the case of pions the agreement is reasonable. The differences
between the calculated curves and the data are not larger than
10\%. In the case of secondary kaons we see that there is a
problem. The spectrum of $K^-$ also is in reasonable agreement
with the model, whereas for $K^+$ disagreement is about 30\%. Even
more serious is the $K^+/K^-$ ratio. Experimentally this ratio at
$y \sim 0$ is about 1.5, however in the model we can not obtain more
than 1.2, the ratio about 1.5 being obtained for $pp$ collisions
at the discussed energy. Most probably it means that the difference
between fragmentation functions of diquarks and/or quarks into $K^+$
and $K^-$ is not large enough in the model. The used fragmentation
functions were taken from \cite{KaPi}. In the cases of $p$ and
$\bar{p}$ the agreement with the data is good enough. The contribution
from string junction diffusion \cite{ACKS}-\cite{AMS} to the proton
spectrum is small at this energy.

\section{\bf Inclusive spectra at high energies}

The Multiple Scattering Theory allows one to obtain some simple and
model independent formulae \cite{PaRa,BSh} coming only from the assumption
that high energy heavy ion collision can be consider as the
superposition of independent nucleon-nucleon collisions. So any serious
disagreement of these predictions with the data can be considered as the
signal for some collective interaction.

The predictions for rapidity spectra of secondaries are more model
dependent. The RHIC experimental data give clear evidences for the
effects of inclusive density saturation, which reduce the inclusive
density about two times in the central (midrapidity) region in
comparison, say, with predictions of \cite{CMT,Sh6} based on the
superposition picture. The observed phenomena can be explained in the
framework of the space--time picture of high energy interactions and
on Gribov's Reggeon diagram technique \cite{RDT}. In high energy
hadron--nucleus collision there exists inelastic screening corrections
\cite{Gri,Grib}. The same inelastic screening should exist in high
energy heavy ion collision. This effect is very small for integrated
cross sections (because many of them are determined by geometry), but it
is very important \cite{CKTr} for the calculations of secondary
multiplicities and inclusive densities.

At not very high energies the heavy ion collisions can be considered
with the help of standard Glauber approximation, and the inclusive
spectrum of any secondary $h$ produced in the central region is
described by the diagram shown in Fig.~5a, i.e. by the contribution
of a single nucleon--nucleon blob. This diagram immediately gives
Eq.~(11) (and Eq.~(12) after accounting the corrections for
finite energy) for heavy ion inclusive cross section. The contributions
of all other diagrams cancel each other due to AGK cutting rules
\cite{AGK}.

\begin{figure}
\begin{center}
\mbox{\psfig{file=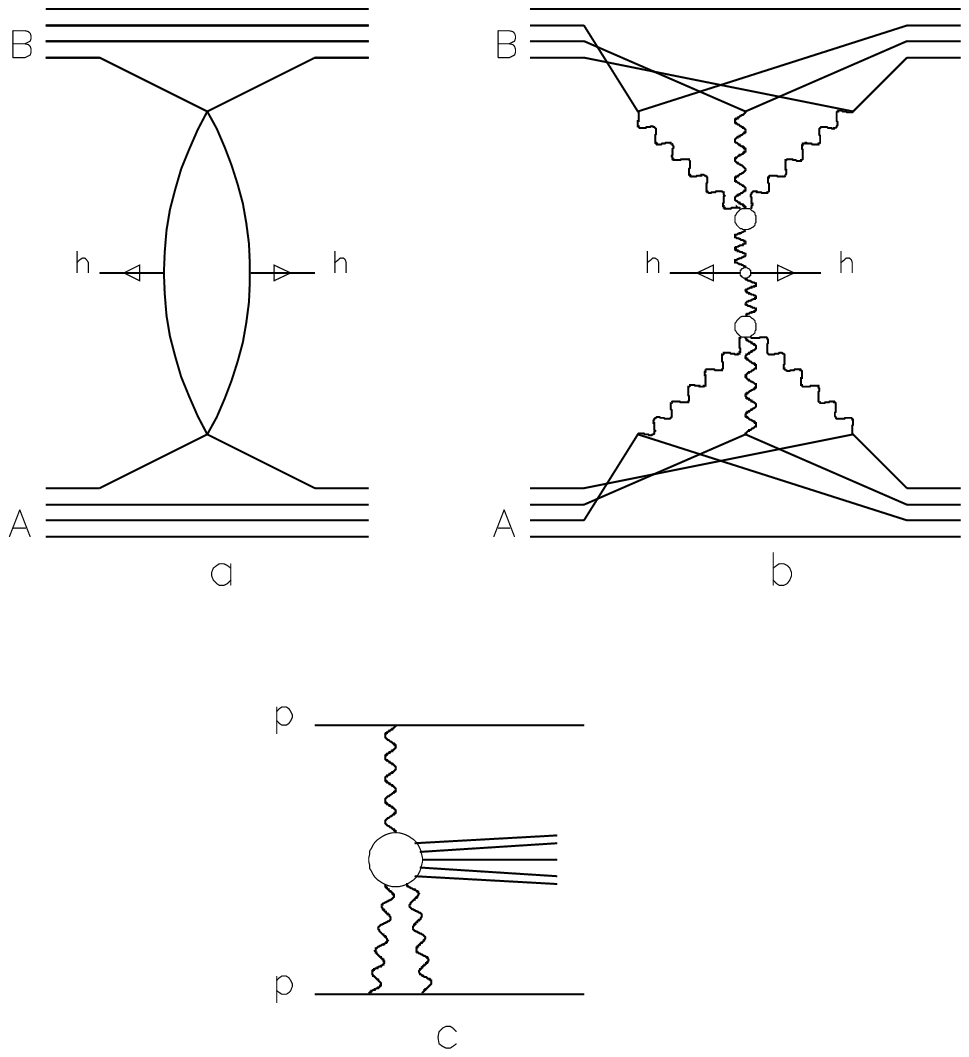,width=.5\textwidth}} \\
Fig.~5. Diagrams for inclusive cross sections for $A-B$ collisions
in Glauber approximation (a) and with accounting for the interactions
of pomerons (shown by wave curves) (b). An example of
inelastic processes of $pp$ interactions which determine one of
the vertices of the multipomeron interactions (c).
\end{center}
\end{figure}

At high energies the new diagrams appear, which include the
interactions of pomerons and correspond to the diffractive production
of a large mass $M$ jet in $pp$ interactions. At low energies the
contribution of such diagrams to nucleus--nucleus interaction is
suppressed by the longitudinal part of nuclear form factor
$G_z(t_{min})$ which is connected with the longitudinal part
of momentum transfer, $q_z$ ($t_{min} = - q_z^2$). Thiz
longitudinal form factor can be written for Gaussian distribution of
nuclear density $\rho(b,z)$ as
\begin{equation}
G_A(t_{min}) = \int \rho_A(b,z) e^{iq_zz} dz \approx
\exp{(R^2_A t_{min}/3)} \;.
\end{equation}
When energy becomes high enough, the value of $t_{min}$ becomes
very small. So the discussed contribution can be significant.

One example of a diagram with pomeron interaction for heavy ion
interaction is shown in Fig.~5b. Contrary to the hadron--nucleus
case, where the inelastic screening is connected with the diffractive
dissociation of projectile particle \cite{Gri,Grib}, the contribution
of such diagram can be estimated from the processes of high mass jet
production in midrapidity region and with two large rapidity gaps, see
for example Fig.~5c. The contribution of the considered diagrams to
inclusive spectrum is suppressed quadratically, by both longitudinal
form factors, $G_A(t_{min})$ and $G_B(t_{min})$. So we can observe
their influence at energies quadratically higher in comparison with
the energies region, where the inelastic screening effects are observed
in hadron--nucleus scattering.

Following to the estimations of \cite{CKTr}, the RHIC energies are of
the needed order of magnitude. The inelastic screening can decrease
\cite{CKTr} the inclusive density in the midrapidity region about two
times at RHIC energies and about three times at LHC energies in
comparison with the calculation without inelastic screening.

However all such estimation are model dependent. The numerical
contribution of all multipomeron diagrams is rather unclear due to
a lot of unknown vertices for the multipomeron interactions, and the
number of multipomeron diagrams is very large. The number of parameters
can be decreased in some model, for example, in \cite{CKTr} the
Schwimmer model \cite{Schw} was used for the numerical estimations.

Another (again model dependent) possibility to estimate the contribution
of the diagrams with Pomeron interaction comes \cite{JUR,BP,JDDSh} from
percolation theory. In this approach we assume that if two or several
pomerons are overlapping, they become a one pomeron. When all Pomerons
are overlapping, the inclusive density is saturated, it reaches its
maximal value at given impact parameter. This approach has one free
parameter - the critical number of pomerons in one squared fermi.
Technically it is more simple to bound the maximal number of pomerons,
$n_{max}$, which can be emitted by one participating nucleon for the
given pair of colliding nuclei. All model calculations become rather
simple because above the critical value every additional pomeron cannot
contribute to the inclusive spectrum.

\begin{figure}
\begin{center}
\vskip -2cm
\mbox{\psfig{file=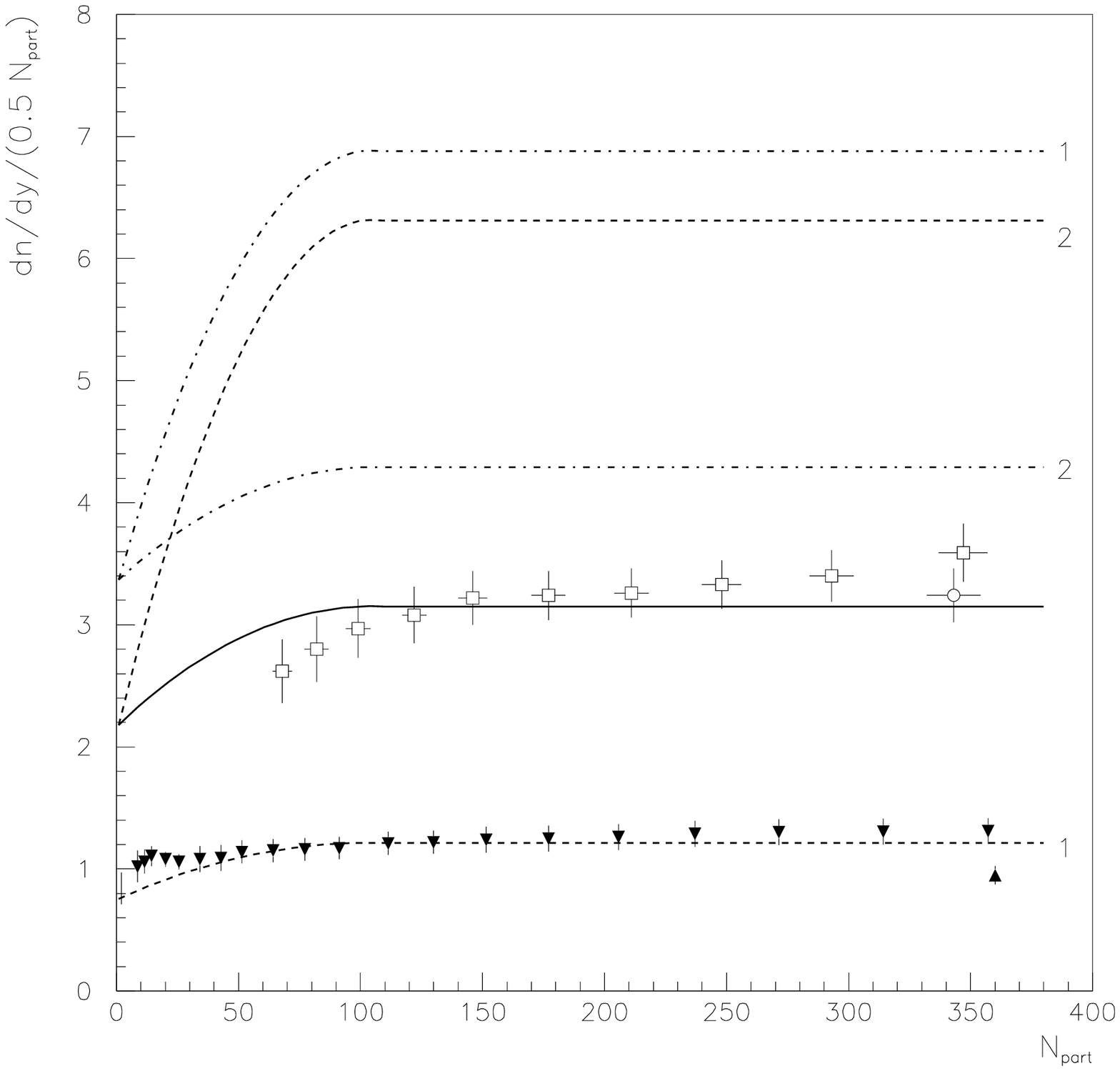,width=.5\textwidth}} \\
Fig.~6. Relative inclusive densities of secondaries for $Pb-Pb$
collisions at $\sqrt{s} = 17.3$ GeV per nucleon (black points,
multiplied by 1/2) and for $Au-Au$ at $\sqrt{s} = 130$ GeV per
nucleon (open points). Dashed curves present the QGSM results without
percolation effects for $\sqrt{s} = 17.3$ GeV (curve 1, multiplied by
1/2, as the points) and for $\sqrt{s} = 130$ GeV (curve 2). The solid
curve shows the results of calculations at $\sqrt{s} = 130$ GeV with
percolation effects ($\langle n_{max} \rangle = 1.67$). Dash--dotted
curves 1 and 2 show the predictions for $Pb-Pb$ collisions at 5.5 TeV
per nucleon with $\langle n_{max} \rangle = 1.67$ and
$\langle n_{max} \rangle = 1$, respectively.
\end{center}
\end{figure}

The results of calculations of the inclusive densities of the produced
charged secondaries per one pair on interacting nucleons
$dn_{ch}/dy/(0.5 N_{part})$, with the help of Eq.~(11) and QGSM are
shown in Fig.~6. We present the CERN SPS experimental data for $Pb-Pb$
collisions at $\sqrt{s} = 17.3$ GeV per nucleon
\cite{SpS}-\cite{Phob} for $\vert y \vert < 1$ and RHIC data for
$Au-Au$ at $\sqrt{s} = 130$ GeV per nucleon \cite{Phob,Phen} (these
data are in agreement with the results of more recent measurements).

One can see that CERN SPS data at $\sqrt{s}$ = 17.3 GeV are
described reasonably (dashed curve 1 in Fig.~6) without any additional
screening corrections, by the same way as the results shown in Fig.~4
for different secondary hadrons. It seems to be natural because the
suppression effects of $t_{min}$, Eq.~(13), discussed above should be
very important at comparatively low energy of CERN SpS.

However, the same calculation at RHIC energy $\sqrt{s} = 130$ GeV per
nucleon gives the relative inclusive density two times larger (dashed
curve 2 in Fig.~6) than the data. The agreement with the data can be
obtained only with suppression of multipomeron contributions, namely
by using $\langle n_{max} \rangle = 1.67$ for every interacting
nucleons. In this case the averaged number of cutted Pomerons emitted
by every interacting nucleon is about 1.2 that is significantly
smaller than the number of cutted Pomerons in $pp$ collisions at the
same energy (the last one is about 1.8).

Here we assume that the value $\langle n_{max} \rangle$ does
not depends on $N_{part}$ in the interval $A/4 < N_{part} < A$, i.e.
between minimum bias and very central collisions \cite{PSh}. This
assumption, as one can see from Fig.~6, is in agreement with the
presented RHIC experimental data with the accuracy 10-20 \%. It is
confirmed (within the same accuracy) by the recent data \cite{Phob3}
that the ratio of charged particle multiplicities in central (where
$N_{part} \approx A$) $Au-Au$ and $Cu-Cu$ collisions in midrapidity
region are both at $\sqrt{s} = 200$ GeV and 62.4 GeV only 15 \% large
that the ratio of $Au$ and $Cu$ atomic weights. However, it is
necessary to note that the calculated inclusive densities in heavy
ion collisions per one pair of participants are about 1.6 times larger
than in $pp$ collisions.

The same value of $\langle n_{max} \rangle = 1.67$ allows us
\cite{JShU} to describe the PHOBOS point at $\sqrt{s} = 56$ GeV
\cite{Phob}. It means, that the saturation effects at $\sqrt{s} =
56$ GeV are of the same order of magnitude as at $\sqrt{s} =$
130 GeV and 200 GeV.

The result of PHOBOS Coll. \cite{Phob2} for central $Au-Au$ collisions
at $\sqrt{s} = 200$ GeV per nucleon gives $d n_{ch}/d \eta = 650 \pm 35$
for $\vert \eta \vert < 1$ that is in agreement with our result
$dn_{ch}/d\eta \approx 600$. The important point is that the
experimental data \cite{NA50} of NA50 Coll. for central $Pb-Pb$
collisions show the increase of the multiplicities of secondaries,
$dn_{ch}/d\eta$, in the energy interval $\sqrt{s}$ = 8.8 GeV -- 17.3 GeV
about two times, from $207 \pm 1 \pm 16$ to $428 \pm 1 \pm 34$.
In the much larger (in logarithmical scale) interval $\sqrt{s}$ =
17.3 GeV -- 200 GeV these multiplicities increase only about 1.5 times.
Probably it means that the inelastic screening (percolation) effects
start to work at energies about $\sqrt{s} \sim 20 - 30$ GeV and they
become very significant at $\sqrt{s} \sim 50$ GeV.

More detailed data are presented in \cite{PHEN1} and in \cite{PHEN2},
where the multiplicities of identified hadrons were measured in $Au-Au$
at $\sqrt{s}$ = 130 GeV and 200 GeV, respectively. The values of
$dn/dy/(0.5N_{part})$ for different secondaries produced in 5 \% of the
most central $Au-Au$ interaction are compared with our QGSM
calculations in the Table. Percolation effects were accounted for, as
it was explained above. The agreement of the QGSM calculations with the
data is on the level of 20 \% that is usual for QGSM (let us note that we
did not input any new parameter). In particular, one can see qualitative
agreement of the calculated results and the data in the energy
dependences of midrapidity multiplicities for secondary pions and kaons,
but rather strange disagreement for secondary protons and antiprotons.

%\newpage

\begin{center}
{\bf Table}
\end{center}
\vspace{15pt}

The values of $dn/dy/(0.5N_{part})$ for different secondaries produced
in 5 \% of the most central $Au-Au$ collisions at $\sqrt{s}$ = 130 GeV
and 200 GeV.

\begin{center}
\vskip 12pt
\begin{tabular}{|c|c|c|c|c|} \hline
Hadron & \multicolumn{2}{c|}{130 GeV} &
\multicolumn{2}{c|}{200 GeV}  \\ \cline{2-5}
& \cite{PHEN1} & QGSM & \cite{PHEN2} & QGSM \\   \hline
$\pi^+$  & $1.59 \pm 0.05$ & 1.82 & $1.63 \pm 0.13$ & 2.00  \\
$\pi^-$  & $1.55 \pm 0.05$ & 1.88 & $1.61 \pm 0.13$ & 2.05  \\
$K^+$   & $0.27 \pm 0.02$ & 0.17 & $0.28 \pm 0.04$ & 0.19 \\
$K^-$   & $0.23 \pm 0.02$ & 0.17 & $0.26 \pm 0.04$ & 0.18 \\
p      & $0.16 \pm 0.01$ & 0.13 & $0.10 \pm 0.01$ & 0.14 \\
$\bar{p}$ & $0.11 \pm 0.01$ & 0.08 & $0.08 \pm 0.01$ & 0.10 \\
\hline
\end{tabular}
\end{center}

\vskip 0.5cm

The results of Dual String Model \cite{DDU,DDU1} calculations with
string fusion \cite{JShU} are in agreement with the results of the
percolation model. The results \cite{APS} of String Fusion Model
\cite{ABP} are also in agreement with presented calculations.

In the case of LHC energy $\sqrt{s} = 5.5$ TeV per nucleon for
$Pb-Pb$ collisions with $\langle n_{max} \rangle = 1.67$ percolation
effect decreases the relative inclusive density about 3 times. This
result is shown by dash-dotted curve 1 in Fig.~6 and it is again
similar to \cite{CKTr} prediction. Let us note that the energy 
difference in calculations with and without percolation effects
is connected mainly with the increase of the total inelastic $NN$
cross section.

However, the percolation effect at LHC energy can be even larger if
the increase of the transverse range of a pomeron with incident energy
will be accounted for \cite{JShU}. The predictions for the relative
inclusive density at LHC energy with $\langle n_{max} \rangle = 1$
(maximal percolation) are shown in Fig.~6 by dash-dotted curve 2. Now
the percolation effect decreases the relative inclusive density about 5
times.

The data \cite{Phob1} on the total number of charged particles
detected within range $-5.4 < \eta < 5.4$, i.e. in the region which
exclude only high-$x_F$ fragmentation and diffraction regions are
presented in Fig.~7.

\begin{figure}
\begin{center}
\vskip -2cm
\mbox{\epsfig{file=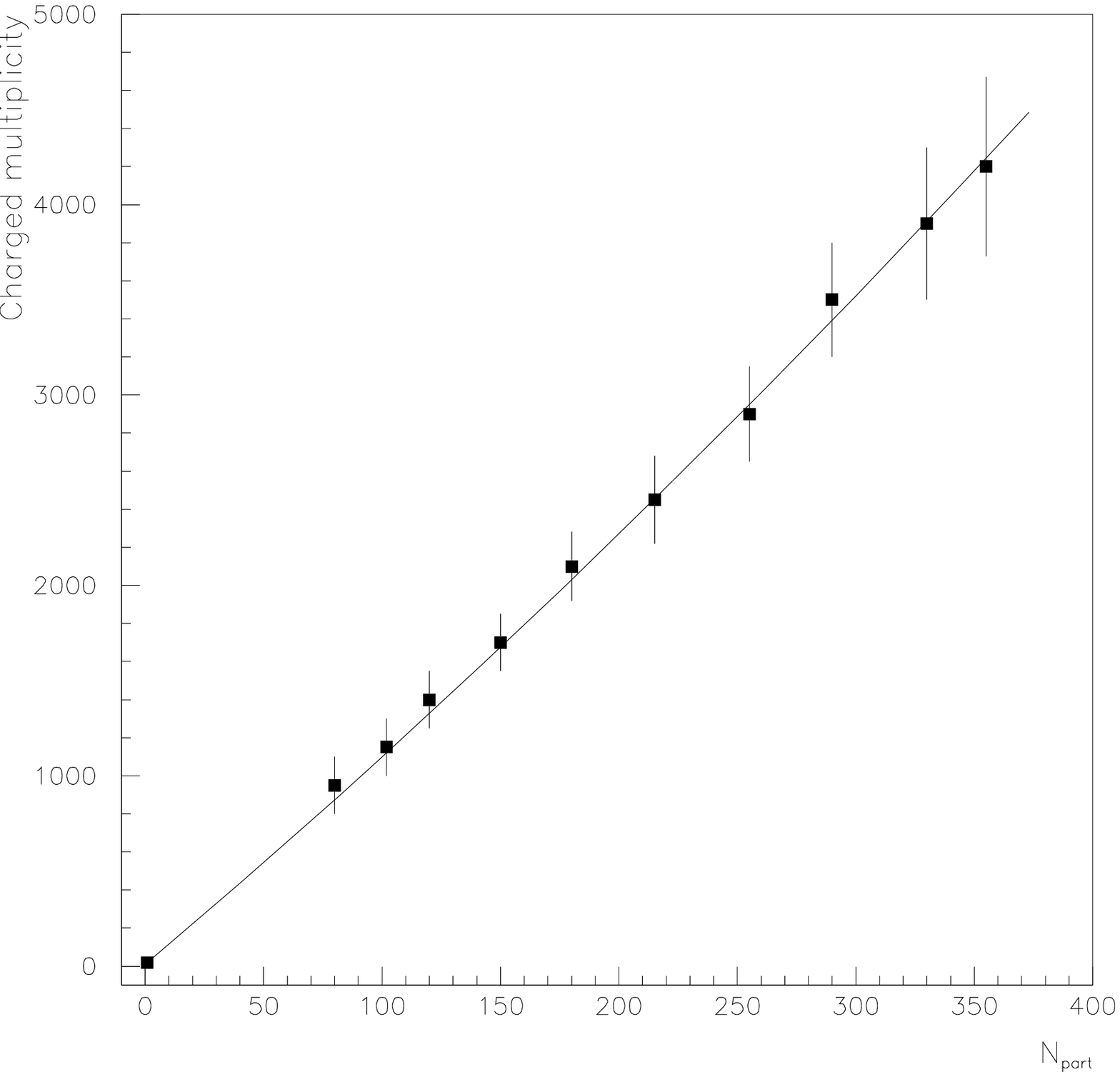,width=.55\textwidth}} \\
Fig.~7. Total number of charged particles detected within range
$-5.4 < \eta < 5.4$ in $Au-Au$ collisions at $\sqrt{s} = 130$ GeV per
nucleon as a function of $N_{part}$ and its description by
the QGSM with percolation effects and $\langle n_{max} \rangle = 1.67$
solid curve.
\end{center}
\end{figure}

These data are in good agreement with the QGSM calculations accounting
for the percolation effects. The calculated dependence are very close
to a straight line that is the direct consequence of Eq.~(11).

\section{Conclusion}

We can see that the QGSM together with Multiple Scattering Theory
can describe on reasonable level the inclusive spectra of different
secondaries produced in heavy ion collisions at not very high
energies. Some disagreement in $K^+$ and $K^-$ yields in Fig.~4
can be corrected by reasonable change of the diquark and quark
fragmentation functions into different kaons.

The data of RHIC and their comparison with CERN SPS data show
numerically large effects coming from the Pomeron (secondary particle)
density saturation. It is the first experimental evidence of so
numerically large effects in high energy physics. The inelastic 
screening effects should be accounted for in calculations of inclusive 
spectra of different secondaries produced in heavy ion collisions at 
RHIC and LHC energies. The processes of baryon number transfer via 
string junction diffusion \cite{ACKS}--\cite{AMS} also should be 
accounted at these energies.

From the comparison of the RHIC data (where effects are large) and
CERN SPS data (where effects are small) we can conclude that the
inelastic screening (saturation effects) become very important for
heavy ion collisions in the energy interval $\sqrt{s_{NN}}$ = 20 -
30 GeV and they, probably, are saturated at $\sqrt{s_{NN}}$ =
60 GeV. At higher energies the growth of total inelastic $NN$ cross
section results in the increase of the number of Pomerons which can be 
screened, so the difference in calculations with and without inelastic
shadowing becomes larger.

We are grateful to N. Armesto, A. Capella, C. Pajares and
R. Ugoccioni for useful discussions. This paper was
supported, in part, by grants RSGSS-1124.2003.2 and PDD (CP)
PST.CLG980287.


\newpage

\begin{thebibliography}{99}

\bibitem{KTM} A. B. Kaidalov, K. A. Ter--Martirosyan, Yad. Fiz. {\bf
39}, 1545 (1984); {\bf 40}, 211 (1984).
\bibitem{KaPi} A. B. Kaidalov, O. I. Piskunova, Yad. Fiz. {\bf 41},
1278 (1985).
\bibitem{2r} A. Capella, U. Sukhatme, C. I. Tan, J. Tran Thanh Van,
Phys. Rep. {\bf 236}, 225 (1994).
\bibitem{CaTran} A. Capella, J. Tran Thanh Van, Z. Phys. C{\bf 10},
249 (1981).
\bibitem{KTMS} A. B. Kaidalov, K. A. Ter-Martirosyan, Yu. M. Shabelski,
Yad. Fiz. {\bf 43}, 1282 (1986).
\bibitem{Sh} Yu. M. Shabelski, Yad. Fiz. {\bf 44}, 186 (1986).
\bibitem{Sh1} Yu. M. Shabelski, Nucl. Phys. Proc. Suppl. {\bf 52B}, 116
(1997).
\bibitem{AGK} V. A. Abramovsky, V. N. Gribov, O. V. Kancheli, Yad. Fiz.
{\bf 18}, 595 (1973).
\bibitem{Kai} A. B. Kaidalov, Sov. J. Nucl. Phys. {\bf 45}, 902 (1987);
Yad. Fiz. {\bf 43}, 1282 (1986).
\bibitem{CM} W. Czyz and L. G. Maximon, Ann. Phys. {\bf 52}, 59 (1969).
\bibitem {Alkh} G. D. Alkhazov et al., Nucl. Phys. {\bf A220}, 365
(1977).
\bibitem{Andr} I. V. Andreev and A. V. Chernov, Yad. Fiz. {\bf 28},
477 (1978).
\bibitem{Bra} M. A. Braun, Yad. Fiz. {\bf 45}, 1625 (1987).
\bibitem{BoKa} K. G. Boreskov and A. B. Kaidalov, Yad. Fiz. {\bf 48},
575 (1988).
\bibitem{E866} L. Ahle et al., Nucl. Phys. {\bf A610}, 139c (1996).
\bibitem{E802} L. Ahle et al., Phys. Rev. {\bf C58}, 3523 (1998).
\bibitem{Sik} F. Sikler et al., NA49 Coll., Nucl.Phys. {\bf A661},45c
(1999).
\bibitem{Kaba} S. Kabana et al., NA52 Coll., Nucl.Phys. {\bf A661},
370c (1999).
\bibitem{CKTr} A. Capella, A. Kaidalov and J. Tran Thanh Van, Heavy Ion
Phys. {\bf 9}, 169 (1999).

\bibitem{RV} G. C. Rossi, G. Veneziano, Nucl. Phys. B{\bf 123}, 507
(1977).
\bibitem{TM} K. A. Ter--Martirosyan, Phys. Lett. {\bf 44B}, 377 (1973).
\bibitem{Sh2} Yu. M. Shabelski, Yad.Fiz. {\bf 26}, 1084 (1977);
Nucl. Phys. {\bf B132}, 491 (1978).
\bibitem{BT} L. Bertocchi and D. Treleani, J. Phys. {\bf G3}, 147 (1977).
\bibitem{Weis} J. Weis, Acta Phys. Polonica {\bf B7}, 85 (1977).
\bibitem{Jar} T. Jaroszewicz et al., Z. Phys. {\bf C1}, 181 (1979).
\bibitem{PSh} C. Pajares and Yu. M. Shabelski, Yad. Fiz. {\bf 63},
980 (2000).
\bibitem{Sh3} Yu. M. Shabelski, Acta Phys. Polonica {\bf B10}, 1049
(1979).
\bibitem{Sh12} Yu. M. Shabelski, Yad. Fiz. {\bf 50}, 239 (1989).
\bibitem{Afan} S. V. Afanasiev et al., NA49 Coll. Phys. Rev. {\bf C66},
054902 (2002).
\bibitem{ACKS} G. H. Arakelyan, A. Capella, A. B. Kaidalov and Yu. M.
Shabelski, Eur. Phys. J. {\bf C26}, 81 (2002); hep-ph/0103337.
\bibitem{SJ1} F. Bopp and Yu. M. Shabelski, Yad. Fiz. {\bf 68}, 2155
(2005); hep-ph/0406158.
\bibitem{SJ2} F. Bopp and Yu. M. Shabelski, Eur. Phys. J. {\bf A28},
237 (2006); hep-ph/0603193.
\bibitem{AMS} G. H. Arakelyan, C. Merino and Yu. M. Shabelski,
Yad. Fiz. {\bf 69}, 911 (2006); hep-ph/0505100.
\bibitem{PaRa} C. Pajares and A. V. Ramallo, Phys. Rev. {\bf C16},
2800 (1985).
\bibitem{BSh} V. M. Braun and Yu. M. Shabelski, Int. J. of Mod. Phys.
{\bf A3}, 2417 (1988).
\bibitem{CMT} A. Capella, C. Merino and J. Tran Thanh Van, Phys. Lett.
{\bf B265} (1991) 415.
\bibitem{Sh6} Yu. M. Shabelski, Z. Phys. {\bf C57}, 409 (1993).
\bibitem{RDT} V. N. Gribov, ZhETF {\bf 53}, 654 (1967).
\bibitem{Gri} V. N. Gribov, ZhETF {\bf 56}, 892 (1968).
\bibitem{Grib} V. N. Gribov, ZhETF {\bf 57}, 1306 (1969).
\bibitem{Schw} A. Schwimmer, Nucl. Phys. {\bf B94}, 445 (1975).
\bibitem{JUR} J. Dias de Deus, R. Ugoccioni and A. Rodrigues,
Phys. Lett. {\bf B458}, 402 (1999); Eur. Phys. J. {\bf C16}, 537 (2000).
\bibitem{BP} M. A. Braun and C. Pajares, Phys. Rev. Lett. {\bf 85},
4864 (2000).
\bibitem{JDDSh} J. Dias de Deus and Yu. M. Shabelski, Eur. Phys. J.
{\bf A20}, 457 (2004).
\bibitem{SpS} NA49 Coll. P. Jones et al., Nucl. Phys. {\bf A610}, 188c
(1996).
\bibitem{SpS2} NA49 Coll. H. Appelshauser et al., Phys. Rev. Lett.
{\bf 82}, 2471 (1999).
\bibitem{SpS1} WA98 Coll. M. M. Aggarwal et al., Eur. Phys. J. {\bf C18},
651 (2001).
\bibitem{Phob} PHOBOS Coll. B. B. Back et al., Phys. Rev. Lett. {\bf 85},
3100 (2000).
\bibitem{Phen} PHENIX Coll. K. Adcox et al., Phys. Rev. Lett. {\bf 86},
3500 (2001).
\bibitem{Phob3} PHOBOS Coll. B. B. Back et al., nucl-ex/0604017.
\bibitem{JShU} J. Dias de Deus, Yu. M. Shabelski and R. Ugoccioni,
hep-ph/0108253.
\bibitem{Phob2} PHOBOS Coll. B. B. Back et al., Phys. Rev. Lett.
{\bf 88}, 022302 (2002); nucl-ex/0108009.
\bibitem{NA50} NA50 Coll. M. C. Abreu et al., Phys. Lett. {\bf B530},
33 (2002).
\bibitem{PHEN1} PHENIX Coll. A. Adcox et al., Phys. Rev. {\bf C69},
024904 (2004); nucl-ex/0307010.
\bibitem{PHEN2} PHENIX Coll., S. S. Adler et al., Phys. Rev. {\bf C69},
034909 (2004); nucl-ex/0307022.
\bibitem{DDU} J. Dias de Deus and R. Ugoccioni, Phys. Lett. {\bf B491},
253 (2000).
\bibitem{DDU1} J. Dias de Deus and R. Ugoccioni, Phys. Lett. {\bf B494},
53 (2000).
\bibitem{APS} N. Armesto, C. Pajares and D. Sousa, Phys. Lett.
{\bf B527}, 92 (2002); hep-ph/0104269.
\bibitem{ABP} N. S. Amelin, M. A. Braun and C. Pajares, Phys. Lett.
{\bf B306}, 312 (1993); Z.Phys. {\bf C63}, 507 (1994).
\bibitem{Phob1} PHOBOS Coll. B. B. Back et al., Phys. Rev. Lett.
{\bf 87}, 102303 (2001); nucl-ex/0106006.

\end{thebibliography}
\end {document}